\documentclass[10pt,twocolumn,amsmath,amssymb,pre]{revtex4}

\usepackage{amsmath}
\usepackage{amssymb}
\usepackage[sort&compress]{natbib}
\usepackage{enumerate}
\usepackage{graphicx}
\usepackage{dcolumn}
\usepackage{bm}
\usepackage{wasysym}
\usepackage{ae,aecompl}

\usepackage[english]{babel}
\usepackage[latin2]{inputenc}
\usepackage[T1]{fontenc}

\def\ccc#1;#2{\left\langle #1 \left\vert #2 \right.\right\rangle}
\def\ev #1{\left\langle #1 \right\rangle}

\begin{document}

\title{Limitations of scaling and universality in stock market data}
\author{J\'anos Kert\'esz}
\altaffiliation[Also at ]{Laboratory of Computational Engineering, Helsinki
University of Technology, Espoo, Finland} 
\author{Zolt\'an Eisler}
\email{eisler@maxwell.phy.bme.hu}
\affiliation{Department of Theoretical Physics, Budapest University of Technology and Economics, Budapest, Hungary}

\begin{abstract}
We present evidence, that if a large enough set of high resolution stock market data is analyzed, certain analogies with physics -- such as scaling and universality -- fail to capture the full complexity of such data. Despite earlier expectations, the mean value per trade, the mean number of trades per minute and the mean trading activity do not show scaling with company capitalization, there is only a non-trivial monotonous dependence. The strength of correlations present in the time series of traded value is found to be non-universal: The Hurst exponent increases logarithmically with capitalization. A similar trend is displayed by intertrade time intervals. This is a clear indication that stylized facts need not be fully universal, but can instead have a well-defined dependence on company size.
\end{abstract}

\maketitle

In the last decade, an increasing number of physicists is becoming devoted to
the study of economic and financial phenomena \cite{evolving, evolving2, kertesz.econophysics}.
One of the reasons for this tendency is that societies or stock markets can be seen
as strongly interacting systems. Since the early $70$'s, physics has developed a wide
range of concepts and models to efficiently treat such topics, these include (fractal and multifractal) scaling, frustrated disordered systems, and far from equilibrium phenomena. To understand how similarly complex patterns arise
from human activity, albeit truly challenging, seems a natural continuation of such efforts.

While a remarkable success has been achieved \cite{bouchaud.book,
stanley.book, mandelbrot.econophysics}, studies in econophysics are
often rooted in possible analogies, even though there are important
differences between physical and financial systems. Despite the
obvious similarities to interacting systems here we would like to
emphasize the discrepancy in the levels of description. For example,
in the case of a physical system undergoing a second order phase
transition, it is natural to assume scaling on profound theoretical
grounds and the (experimental or theoretical) determination of, e.g.,
the critical exponents is a fully justified undertaking. There is no
similar theoretical basis for the financial market whatsoever,
therefore in this case the assumption of power laws should be
considered only as one possible way of fitting fat tailed
distributions \cite{gopi.inversecube, lux.paretian}. Also, the
reference to universality should not be plausible as the robustness of
\emph{qualitative} features -- like the fat tail of the distributions
-- is a much weaker property. While we fully acknowledge the process
of understanding based on analogies as an important method of
scientific progress, we emphasize that special care has to be taken in
cases where the theoretical support is sparse.

The aim of this paper is to summarize some recent advances that help
to understand these fundamental differences. We present evidence, that
the size of companies strongly affects the characteristics of trading
activity of their stocks, in a way which is incompatible with the
popular assumption of universality in trading dynamics. Instead,
certain stylized facts have a well-defined dependence on company
capitalization. Therefore, e.g., averaging distributions over
companies with very different capitalization is questionable.

The paper is organized as follows. Section \ref{sec:intro} introduces
the notations and data that were used. Section \ref{sec:cap} shows
that various measures of trading activity depend on capitalization in
a non-trivial way. In Sec. \ref{sec:correl}, we analyze the
correlations present in traded value time series, and find that the
Hurst exponent increases with the mean traded value per minute
logarithmically.  Section \ref{sec:itt} deals with a similar
size-dependence of correlations present in the time intervals between
trades. Finally, Section \ref{sec:conc} concludes.

\section{Notations and data}
\label{sec:intro}
For time windows of size $\Delta t$, let us write the total traded value
(activity, flow) of the $i$th stock at time $t$ as 
\begin{equation}
f_i^{\Delta t}(t) = \sum_{n, t_i(n)\in [t, t+\Delta t]} V_i(n),
\label{eq:flow}
\end{equation} 
where $t_i(n)$ is the time of the $n$-th transaction of the $i$-th stock.
This corresponds to the coarse-graining of the individual
events, or the so-called tick-by-tick data. $V_i(n)$ is the value traded in transaction $n$, and
it can be calculated as the product of the price $p$ and the traded volume of stocks $\tilde V$,
\begin{equation}
V_i(n) = p_i(n) \tilde V_i(n).
\label{eq:v}
\end{equation}

Price does not change very much from trade to trade, so the dominant
factor in the fluctuations and the statistical properties of $f$ is
given by the variation of the number of stocks exchanged in the
transactions, $\tilde V$. Price serves as a conversion factor to a
common unit (US dollars), and it makes the comparison of stocks
possible, while also automatically corrects the data for stock
splits. The statistical properties (normalized distribution,
correlations, etc.) are otherwise practically indistinguishable
between traded volume and traded value.

We used empirical data from the TAQ database \cite{taq1993-2003} which records all transactions of the New York
Stock Exchange and NASDAQ for the years $1993-2003$.

Finally, note that throughout the paper we use $10$-base logarithms.

\section{Capitalization affects basic measures of trading activity}
\label{sec:cap}

Most previous studies are restricted to an analysis of the stocks of large
companies. These are traded frequently, and so price and
returns are well defined even on the time scale of a few seconds.
Nevertheless, other quantities regarding the activity of trading, such as traded value and volume or the number of
trades can be defined, even for those stocks where they are zero for
most of the time. In this section we extend the study of Zumbach
\cite{zumbach} which concerned the $100$ large companies included in London Stock Exchange's
FTSE-100 market index. This set spans about two orders of magnitude in capitalization.
Instead, we analyze the $3347$ stocks\footnote{Note that many minor stocks do not represent
actual companies, only different class stocks of a larger firm.} that were traded continuously at NYSE for
the year $2000$. This gives us a substantially larger range of capitalization, approximately $10^6\dots 6\cdot
10^{11}$ USD.

\begin{figure*}[ht]
\centerline{\hbox{\includegraphics[width=183pt]{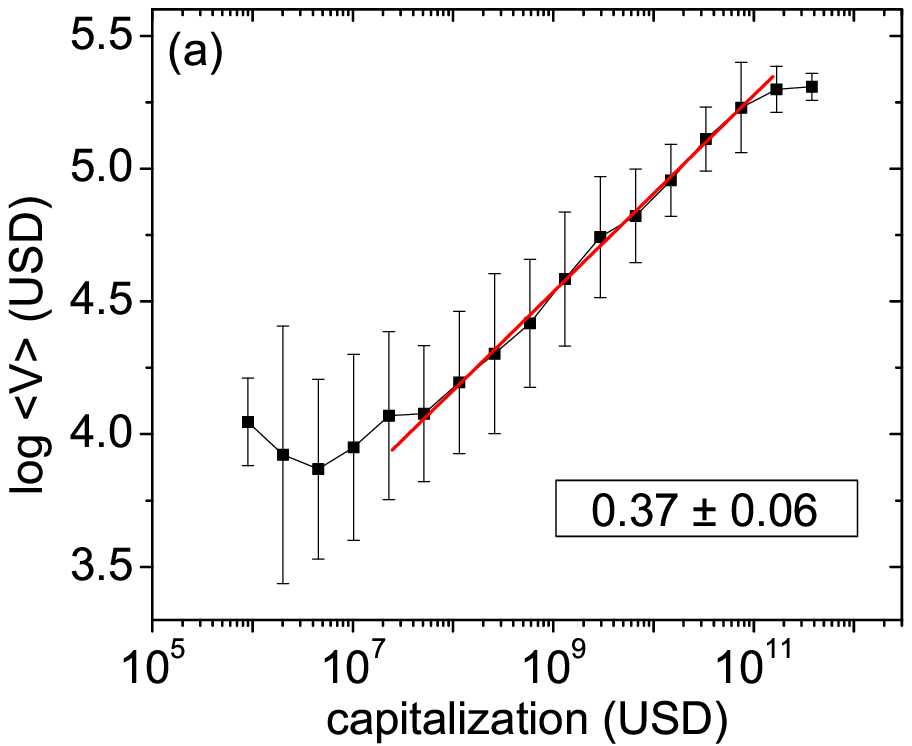}\hskip1mm\hbox{\includegraphics[width=178pt]{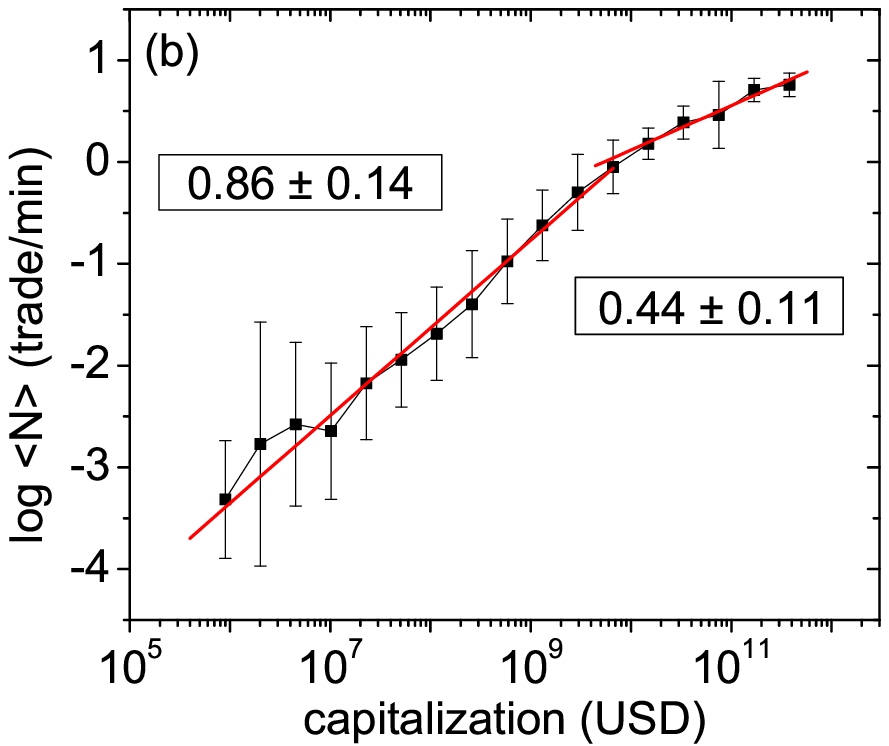}}}}
\centerline{\hbox{\includegraphics[width=195pt]{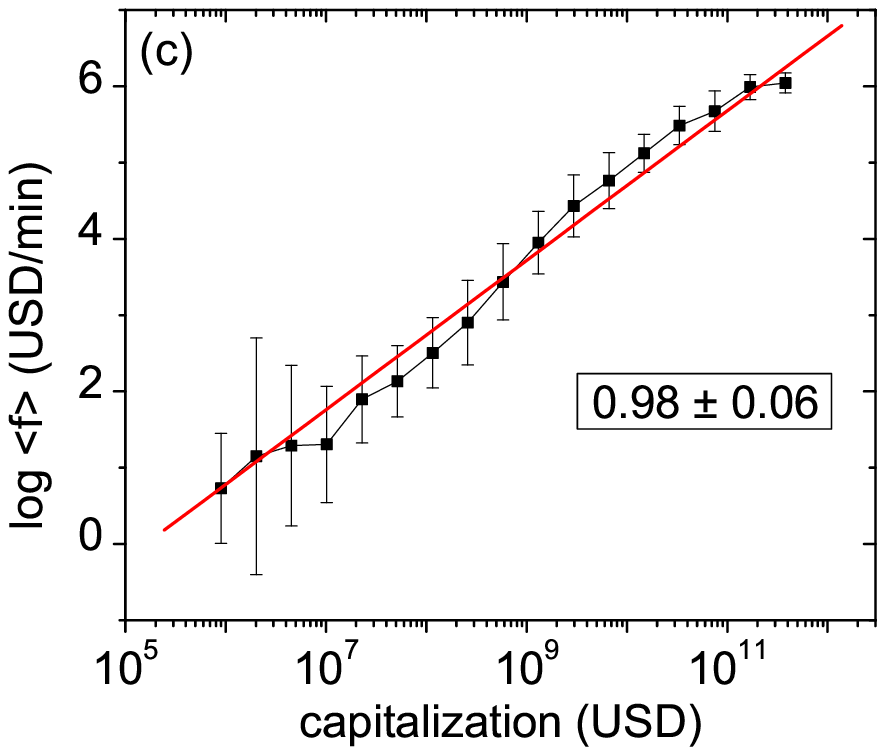}\hskip-6mm\hbox{\includegraphics[width=185pt]{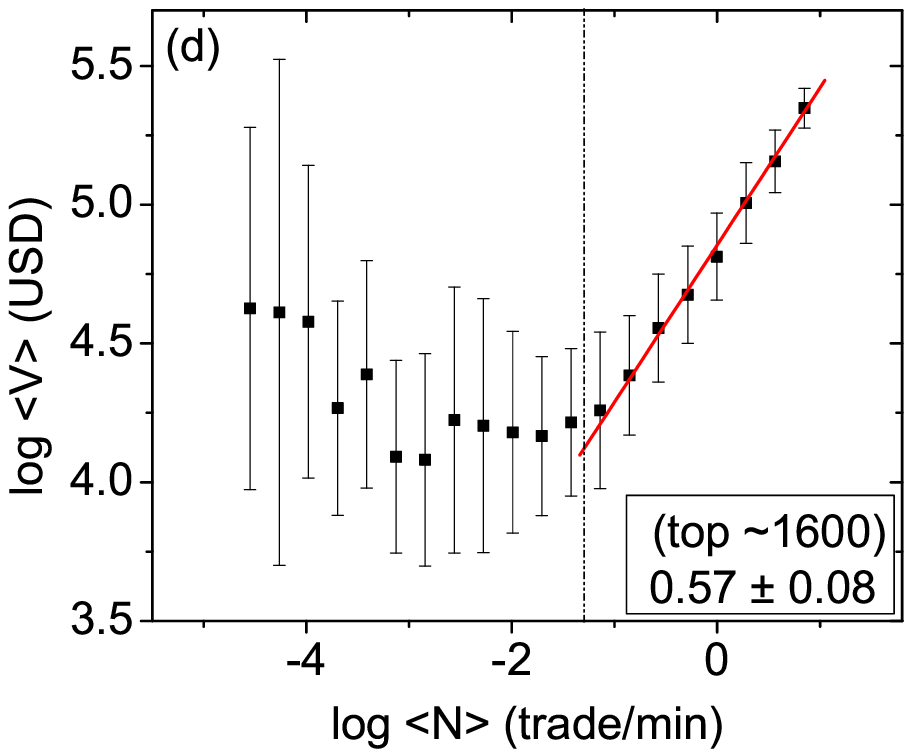}}}}
\caption{{\bf (a)-(c)} Capitalization dependence of certain measures of trading
activity in the year $2000$. The functions are monotonously increasing and
can be piecewise approximated by power laws as indicated. All
three tendencies break down for large capitalizations. {\bf (a)}
Mean value per trade $\ev{V}$ in USD. The fitted slope corresponds to
the regime $5\cdot 10^7<C<7.5\cdot 10^{10}$ in USD. {\bf (b)} Mean
number of trades per minute $\ev{N}$. The slope on the left is from a
fit to $C<4.5\cdot 10^9$ USD, while the one on the right is for
$C>4.5\cdot 10^9$ USD. {\bf (c)} Mean trading activity (exchanged
value per minute) $\ev{f}$ in USD. The plots include
$3347$ stocks that were continuously available at NYSE during $2000$.
{\bf (d)} Plot of mean value per trade $\ev{V}$ versus mean number of
trades per minute $\ev{N}$ for the year $2000$ of NYSE. For smaller
stocks there is no clear tendency. For the top $\sim1600$ companies
($\ev{N} > 0.05$ trades/min), however, there is scaling with an
exponent $\beta = 0.57 \pm 0.08$.}
\label{fig:capdep}
\end{figure*}

Following Ref. \cite{zumbach}, in order to quantify how the value of the capitalization $C_i$ of a company is reflected in the trading activity of its stock, we plotted the mean value per trade $\ev{V_i}$, mean number of trades per minute $\ev{N_i}$ and mean activity (traded value per minute) $\ev{f_i}$ versus capitalization in
Fig. \ref{fig:capdep}. Ref. \cite{zumbach} found that all three quantities have power law dependence on $C_i$, however, this simple ansatz does not seem to work for our extended range of stocks. While mean trading activity can be -- to a reasonable quality -- approximated as $\ev{f_i} \propto C_i^{0.98\pm0.06}$, neither $\ev{V}$ nor $\ev{N}$ can be fitted by a single power law in the whole range of capitalization. Nevertheless, there is an unsurprising monotonous dependence: higher capitalized stocks are traded more intensively.

One can gain further insight from Fig. \ref{fig:capdep}(d), which eliminates the capitalization variable, and shows
$\ev{V}$ versus $\ev{N}$. For the largest $1600$ stocks we find the scaling relation
\begin{equation}	
\ev{V_i} \propto \ev{N_i}^\beta ,
\label{eq:vvsn}
\end{equation}
with $\beta = 0.57 \pm 0.09$. The estimate based on the results of Zumbach \cite{zumbach} for the stocks in London's FTSE-100, is $\beta \approx 1$, while Ref. \cite{eisler.unified} finds $\beta = 0.22 \pm 0.04$ for NASDAQ. The regime of smaller stocks shows no clear tendency. 

One possible interpretation of the effect is the following. Smaller stocks are exchanged rarely,
but there must exist a smallest exchanged value that is still profitable
to use due to transaction costs, $\ev{V}$ cannot decrease indefinitely. On the other hand, once a stock is
exchanged more often (the change happens at about $\ev{N} =
0.05$ trades/min), it is no more traded in this minimal profitable unit. With
more intensive trading, trades "stick together", liquidity allows
the exchange of larger packages. This increase is clear, but not very large,
up to one order of magnitude. Although increasing
package sizes reduce transaction costs, price impact
\cite{gabaix.powerlaw, plerou.powerlaw, farmer.powerlaw,
farmer.whatreally} increases, and profits will decrease again. The balance between
these two effects can determine package sizes and may play a role in the formation of
\eqref{eq:vvsn}.

\section{Non-universal correlations of traded value}
\label{sec:correl}
Scaling methods \cite{vicsek.book, dfa.intro, dfa} have long been used
to characterize stock market time series, including prices and trading volumes
\cite{bouchaud.book, stanley.book}. In particular,
the Hurst exponent $H(i)$ is often calculated. For the traded value time series $f_i^{\Delta t}(t)$ of stock $i$, it can be defined as
\begin{equation}
\label{eq:hurst}
\sigma_i^2(\Delta t) = \ev{\left (f_i^{\Delta t}(t)-\ev{f_i^{\Delta t}(t)} \right )^2}\propto\Delta t^{2H(i)},
\end{equation}
where $\ev{\cdot}$ denotes time averaging with respect to $t$. The signal is said to be correlated (persistent)
when $H>0.5$, uncorrelated when $H=0.5$, and anticorrelated (antipersistent) for $H<0.5$. It is not a trivial fact, but several recent papers \cite{eisler.sizematters, queiros.volume} point out that the variance on the left hand side exists for any stock's traded value and any time scale $\Delta t$. Therefore, we carried out measurements of $H$ on all $2647$ stocks that were continuously traded on NYSE in the period $2000-2002$. We investigated separately the $4039$ stocks that were traded at NASDAQ for the same period. 

\begin{figure}[htb]
\centerline{\includegraphics[height=195pt]{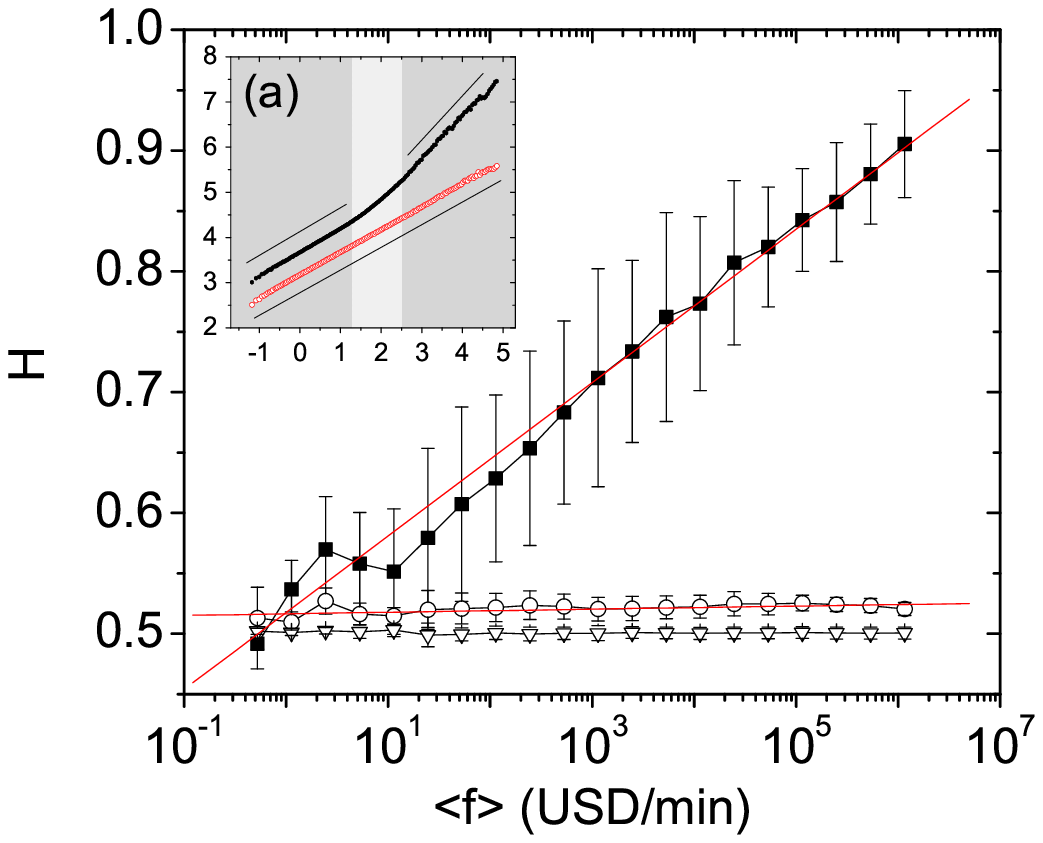}}
\centerline{\includegraphics[height=195pt]{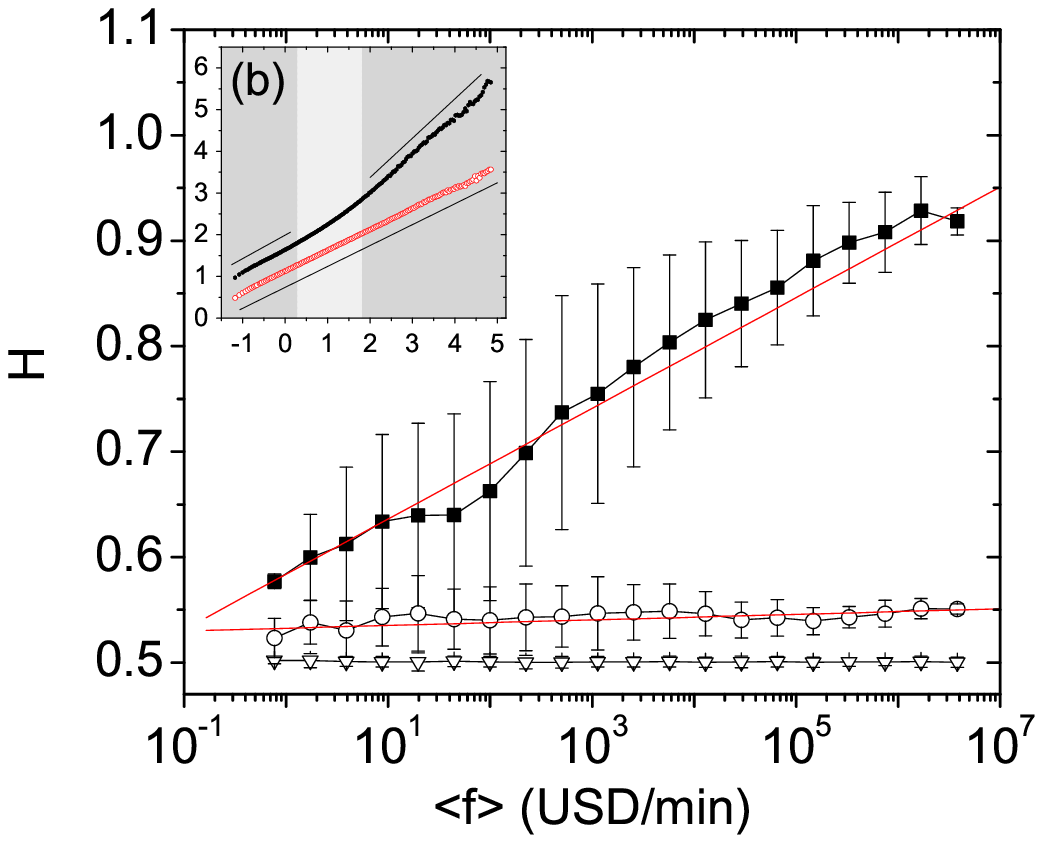}}
\caption{Behavior of the Hurst exponents $H(i)$ for the
period $2000-2002$, and two markets ({\bf (a)} NYSE, {\bf (b)} NASDAQ). For short time windows ($\Circle$), all signals
are nearly uncorrelated, $H(i)\approx 0.51 - 0.52$, regardless of stock market.
The fitted slopes are $\gamma_{NYSE}(\Delta t < \mathrm{20\space min})=0.001\pm 0.002$, and
$\gamma_{NASDAQ}(\Delta t < \mathrm{2\space min})=0.003\pm 0.002$.
For larger time windows ($\blacksquare$), the strength of correlations
depends logarithmically on the mean trading activity of the stock,
$\gamma_{NYSE}(\Delta t > \mathrm{300\space min})=0.06\pm 0.01$ and
$\gamma_{NASDAQ}(\Delta t > \mathrm{60\space min})=0.05\pm 0.01$. 
Shuffled data ($\bigtriangledown$) display no correlations, thus $H_{shuff}(i) = 0.5$.
{\it Insets:} The $\log \sigma$-$\log \Delta t$ scaling plots ($\blacksquare$) for two example stocks,
GE (NYSE) and DELL (NASDAQ). The darker shaded intervals have well-defined Hurst exponents,
the crossover is indicated with a lighter background. Results for shuffled time series ($\Circle$) were
shifted vertically for better visibility.}
\label{fig:hurst}
\end{figure}

We find, that stock market activity has a much richer
behavior, than simply all stocks having Hurst exponents statistically
distributed around an average value, as assumed in 
Ref. \cite{gopi.volume}. Instead, there is a crossover
\cite{eisler.sizematters, ivanov.itt, ivanov.unpublished} between two
types of behavior around the time scale of a few hours to $1$ trading day.
An essentially uncorrelated regime was found when
$\Delta t < 20$ min for NYSE and $\Delta t < 2$ min for NASDAQ, while
the time series of larger companies become strongly correlated when
$\Delta t > 300$ min for NYSE and $\Delta t > 60$ min for NASDAQ. As a
reference, we also calculated the Hurst exponents $H_{shuff}(i)$ of
the shuffled time series. The results are plotted in
Fig. \ref{fig:hurst}.

One can see, that for shorter time windows, correlations are absent in both
markets, $H(i)\approx0.51-0.53$. For windows longer than a trading day, however,
while small $\ev{f}$ stocks
again display only very weak correlations, larger ones show up to $H\approx
0.9$. Furthermore, there is a distinct logarithmic trend in the data:
\begin{equation}
H(i) = H^* + \gamma\log\ev{f_i},
\label{eq:hurst_scaling}
\end{equation}
with $\gamma(\Delta t > 300min) = 0.06\pm0.01$ for NYSE and
$\gamma(\Delta t > 60min) = 0.05\pm0.01$
for NASDAQ. This result can be predicted by a general framework based on a new type of scaling law \cite{eisler.non-universality, eisler.unified}. Shorter time scales correspond to the special case $\gamma = 0$,
there is no systematic trend in $H$. After shuffling the time series, as expected,
they become uncorrelated and show $H_{shuff}(i)\approx 0.5$ at all time scales and without significant
dependence on $\ev{f_i}$.

It is to be emphasized, that the crossover is not simply between uncorrelated
and correlated regimes. It is instead between homogeneous (all stocks
show $H(i)\approx H_1$, $\gamma = 0$) and inhomogeneous ($\gamma > 0$)
behavior. One finds $H_1 \approx 0.5$, but very small
$\ev{f}$ stocks do not depart much from this value even for large time
windows. This is a clear relation to company size, as $\ev{f}$
is a monotonously growing function of company capitalization
(see Sec. \ref{sec:cap} and Ref. \cite{eisler.sizematters}).

Dependence of the effect on $\ev{f}$ is in fact a
dependence on company size. This is a direct evidence of non-universality. The trading mechanism that governs the marketplace depends strongly on the stock that is traded. In a physical sense, there are no universality classes \cite{reichl} comprising a given group of stocks and characterized by a set of stylized facts, such as Hurst exponents. Instead, there is a continuous spectrum of company sizes and the stylized facts may depend \emph{continuously} on company size/capitalization.

Systematic dependence of the exponent of the power spectrum of the
number of trades on capitalization was previously reported in
Ref. \cite{bonanno.dynsec}, based on the study of $88$ stocks.  That
quantity is closely related to the Hurst exponent of the respective time series (see Ref. \cite{ivanov.itt}).
Direct analysis finds a strong, monotonous increase of
the Hurst exponent of $N$ with growing $\ev{N}$, but no such clear logarithmic
trend as Eq. \eqref{eq:hurst_scaling}.

\section{Non-universal correlations of intertrade times}
\label{sec:itt}

To strengthen the arguments of Sec. \ref{sec:correl}, we carried out a
a similar analysis of the intertrade interval series $T_i(n=1\dots N_i-1)$,
defined as the time spacings between the
$n$'th and $n+1$'th trade. $N_i$ is the total number of trades for
stock $i$ during the period under study.

Previously, Ref. \cite{ivanov.itt} used $30$ stocks from the TAQ
database for the period $1993-1996$ and proposed that $H_T$ has the 
universal value $0.94 \pm 0.05$.

We analyzed the same database, but included a large number of stocks with very
different capitalizations. First it has to be noted that the mean intertrade interval has
decreased drastically over the years. In this sense the stock market
cannot be considered stationary for periods much longer than one
year. We analyzed the two year period $1994-1995$ (part of that used in
Ref. \cite{ivanov.itt}) and separately the single year $2000$. We used
all stocks in the TAQ database with $\ev{T} < 10^5$ sec, a total of
$3924$ and $4044$ stocks, respectively.

The Hurst exponents for the time series $T_i$ can written, analogously to Eq. \eqref{eq:hurst}, as
\begin{equation}
\label{eq:hurstittdef}
\sigma_i^2(N) = \ev{\left (\sum_{n=1}^N T_i(n)-\ev{\sum_{n=1}^N T_i(n)} \right )^2}\propto N^{2H_T(i)},
\end{equation}
where the series is not defined in time, but instead on a tick-by-tick basis, indexed by the number of transactions.

\begin{figure}[tb]
\centerline{\includegraphics[height=190pt]{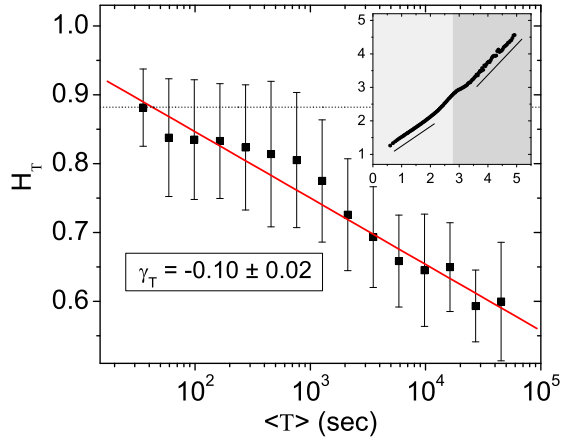}}
\caption{Hurst exponents of $T_i$ for windows greater
than $1$ day, plotted versus the mean intertrade time
$\ev{T_i}$. Stocks that are traded less frequently, show markedly
weaker persistence of $T$ for time scales longer than $1$ day. The
dotted horizontal line serves as a reference. We used stocks with
$\ev{T} < 10^5$ sec, the sample period was $1994-1995$. The inset shows the two
regimes of correlation strength for the single stock General Electric (GE) on
a log-log plot of $\sigma(N)$ versus $N$. The slopes corresponding to Hurst exponents are
$0.6$ and $0.89$.}
\label{fig:ITT}
\end{figure}

The data show a crossover, similar to that for the traded value $f$, from a lower to a higher value of $H_T(i)$ when the window size is approximately the daily mean number of trades (for an example, see the inset of Fig. \ref{fig:ITT}).
For the restricted set studied in Ref. \cite{ivanov.itt}, the value $H_T\approx 0.94\pm0.05$
was suggested for window sizes above the crossover.

Similarly to the case of traded value Hurst exponents analyzed in Section \ref{sec:correl}, the
inclusion of more stocks\footnote{For a reliable calculation of Hurst
exponents, we had to discard those stocks that had less than $\ev{N} <
10^{-3}$ trades/min for $1994-1995$ and $\ev{N} < 2\cdot 10^{-3}$
trades/min for $2000$. This filtering leaves $3519$ and $3775$ stocks,
respectively.} reveals the underlying systematic
non-universality. Again, less frequently traded stocks appear to have
weaker autocorrelations as $H_T$ decreases monotonously with growing
$\ev{T}$. One can fit an approximate logarithmic law \footnote{As
intertrade intervals are closely related to the number of trades per
minute $N(t)$, it is not surprising to find the similar tendency for
that quantity \cite{bonanno.dynsec}.}$\null^,$\footnote{Note that for
window sizes smaller than the daily mean number of trades, intertrade
times are only weakly correlated and the Hurst exponent is nearly
independent of $\ev{T}$. This is analogous to what was seen for traded
value records in Sec. \ref{sec:correl}.} to characterize the trend:
\begin{equation}
H_T = H_T^*+\gamma_T\log\ev{T},
\label{eq:hurstitt}
\end{equation}
where $\gamma_T = -0.10\pm 0.02$ for the period $1994-1995$ (see
Fig. \ref{fig:ITT}) and $\gamma_T = -0.08 \pm 0.02$ for the year
$2000$ \cite{uponrequest}.

In their recent preprint, Yuen and Ivanov \cite{ivanov.unpublished} independently show a tendency similar to Eq. \eqref{eq:hurstitt} for intertrade times of NYSE and NASDAQ in a different set of stocks.

\section{Conclusions}
\label{sec:conc}

In this paper we have summarized a few recent advances in
understanding the role of company size in trading dynamics. We
revisited a number of previous studies of stock market data and found
that the extension of the range of capitalization of the studied firms
reveals a new aspect of stylized facts: The characteristics of trading
display a fundamental dependence on capitalization.

We have shown that trading activity $\ev{f}$, the number of trades per
minute $\ev{N}$ and the mean size of transactions $\ev{V}$ display
non-trivial, monotonous dependence on company capitalization, which
cannot be described by a simple power law.  On the other hand, for
moderate to large companies, a power law gives an acceptable fit for
the dependence of the mean transaction size on the trading frequency.

The Hurst exponents for the variance of traded value/intertrade times
can be defined and they depend logarithmically on the mean trading
activity $\ev{f}$/mean intertrade time $\ev{T}$.

These findings imply that special care must be taken when the concepts
of scaling and universality are applied to financial processes. For
the modeling of stock market processes, one should always consider
that many characteristic quantities depend strongly on the
capitalization. The introduction of such models seems a real challenge
at present.

\section{Acknowledgement}
The authors thank Gy\"orgy Andor for his support with the data. JK is member
of the Center for
Applied Mathematics and Computational Physics, BME; furthermore, he is
grateful for the hospitality of Dietrich Wolf (Duisburg) and of the
Humboldt Foundation. Support by OTKA T049238 is acknowledged.

\bibliographystyle{unsrt}
\bibliography{sizematproc3}

\end{document}